\documentstyle[preprint,aps,prb]{revtex}

\begin{document}

\draft

\def\beq{\begin{equation}}
\def\eeq{\end{equation} }
\def\beqn{\begin{eqnarray}}
\def\eeqn{\end{eqnarray} }

\def\bnabla{\mbox{\boldmath $\nabla$}}
\def\bsigma{\mbox{\boldmath $\sigma$}}
\def\bR{{\bf R}}

\title{ Unbiased estimators in Quantum Monte Carlo
methods: application to liquid $^{\bf 4}$He}

\author{J. Casulleras and J. Boronat}
\address{Departament de
F\'{\i}sica i Enginyeria Nuclear, Campus Nord B4--B5, \protect\\
Universitat Polit\`ecnica de Catalunya, E--08028 Barcelona,
Spain}

%\date{ }

\maketitle

\begin{abstract}

A Monte Carlo algorithm for computing quantum mechanical
expectation values of coordinate operators in many body problems
is presented.  The algorithm, that relies on the forward walking
method, fits naturally in a Green's Function Monte Carlo
calculation, i.e., it does not require side walks or a bilinear
sampling method. Our method evidences stability regions large
enough to accurately sample unbiased pure expectation values.
The proposed algorithm yields accurate results when it is applied
to test problems as the hydrogen atom and the hydrogen molecule.
An excellent description of several properties of a fully many
body problem as liquid $^4$He at zero temperature is achieved.

\end{abstract}


\narrowtext

\section{Introduction}

Quantum Monte Carlo (QMC) methods have become an invaluable tool
in the study of many--body systems over the last decades.  Among
them, the Green's Function Monte Carlo (GFMC) method
\cite{CK79,SK84,Gu88} has been extensively applied to the
calculation of ground--state properties of small molecules and
quantum liquids and solids at zero temperature.  Within the GFMC
techniques one can distinguish between the domain GFMC,
\cite{CK79} which stochastically constructs the Green's function,
and the Diffusion Monte Carlo (DMC) method \cite{RC82} based on a
short--time approximation for the Green's function.  We will
focus our discussion on the DMC method but the algorithm we
present here for the evaluation of pure estimators can also be
easily incorporated in a domain GFMC program.

The DMC method solves the Schr\"odinger equation in imaginary
time for the function $f(\bR,t)=\psi(\bR) \Psi(\bR,t)$,
\beq
-\frac{\partial f(\bR,t)}{\partial t}= -\frac{1}{2} \,
\bnabla_{\bR}^2 f(\bR,t) +
\frac{1}{2} \, \bnabla_{\bR}( {\bf F}(\bR) f(\bR,t) ) +(E_L(\bR)-E) \,
f(\bR,t) \ ,
\label{dmc}
\eeq
being $\Psi(\bR,t)$ the wave function of the system and
$\psi(\bR)$ a trial function used for importance sampling.  In
Eq.  (\ref{dmc}), which is written in atomic units,
$E_L=\psi(\bR)^{-1} H \psi(\bR)$ is the local energy and ${\bf
F}(\bR)=2
\psi(\bR)^{-1} \bnabla_\bR \psi(\bR)$ is the so--called quantum force;
$\bR$ stands for a $3N$--coordinate vector and $E$ is an
arbitrary energy shift.  The Schr\"odinger equation for
$f(\bR,t)$ (\ref{dmc}) presents in the right--hand side three
terms that are associated, by analogy to classical equations, to
diffusion, drift and branching processes, respectively.  The
asymptotic solution of Eq.  (\ref{dmc}), for any value $E$ close
to the energy of the ground state and for long times ($t
\rightarrow \infty$), gives the ground--state wave function
$\Phi_0(\bR)$ provided that there is a nonzero overlap between
$\Psi(\bR,t=0)$ and $\Phi_0(\bR)$.  The formal solution of Eq.
(\ref{dmc}) is
\beq
f(\bR^{\prime},t+\Delta t)=\int d \bR \ G(\bR^{\prime},\bR,
\Delta t) f(\bR,t) \ ,
\label{green}
\eeq
where the Green's function $G(\bR^{\prime},\bR,\Delta t)$ gives
the probability of transition from $\bR$ to $\bR^{\prime}$ in a
time interval $\Delta t$.  The DMC method solves stochastically
Eq.  (\ref{green}) assuming reasonable approximations for the
Green's function when $\Delta t
\rightarrow 0$. \cite{RC82,Chin}  After an iterative process, the
asymptotic solution $f(\bR,t \rightarrow \infty)=\psi(\bR)
\Phi_0(\bR)$ is finally obtained.

The direct calculation of the expectation value of an operator
$A(\bR)$ from the asymptotic function $f(\bR,t \rightarrow
\infty)$ corresponds to the mixed estimator
\beq
\langle
A(\bR) \rangle_m=\frac{\langle \psi(\bR) \,| \, A(\bR) \, |
\, \Phi_0(\bR) \rangle} {\langle \psi(\bR) \, | \, \Phi_0(\bR)
\rangle} \ .
\label{mixed}
\eeq
It gives an exact result only when $A$ is the Hamiltonian $H$ or
commutes with $H$.  Among the different methods to calculate
expectation values for operators that do not commute with $H$,
the extrapolation method \cite{CK79,WC79} is the most widely
used.  Following this method, which has been extensively applied
in QMC calculations, one has an approximation to the ``pure"
(exact) value,
\beq
\langle A(\bR)
\rangle_p=
\frac{\langle \Phi_0(\bR) \, | \, A(\bR) \, | \, \Phi_0(\bR) \rangle}
{\langle \Phi_0(\bR) \, | \, \Phi_0(\bR)  \rangle} \ ,
\label{pure}
\eeq
by means of a linear extrapolation
\beq
\langle A(\bR) \rangle_e=2 \, \langle A(\bR)
\rangle_m - \langle A(\bR) \rangle _v \ ,
\label{extrapol}
\eeq
where
\beq
\langle A(\bR) \rangle_v=
\frac{\langle \psi(\bR) \, | \, A(\bR) \, | \, \psi(\bR) \rangle}
{\langle \psi(\bR) \, | \, \psi(\bR)  \rangle} \ ,
\label{variatio}
\eeq
is the variational estimator of $A(\bR)$.

The accuracy of the extrapolation method is closely related to
the trial wave function used for importance sampling.
Furthermore, in spite of using accurate trial wave functions, the
extrapolated estimator is always biased in a quantity difficult
to assess. In order to overcome these important restrictions,
several algorithms have been proposed in the last years. In the
approach of Zhang and Kalos \cite{ZK93} a bilinear sampling
method is used.  In this scheme, the system is doubled and the
random walks take place in an enlarged configuration space.
Other approaches  are based on the estimation of the quotient
$(\Phi_0/\,\psi)$ from the asymptotic offspring coming from the
branching term. \cite{LK74} In this line, Barnett {\it et al.}
\cite{BR91,RN86} and Runge and Runge \cite{Runge} have
constructed tagging algorithms to properly account for the
asymptotic number of descendants.  Although the later scheme has
provided satisfactory results in some specific cases, the large
fluctuations observed in the asymptotic offsprings, and therefore
in the corresponding weights $(\Phi_0/\,\psi)$, have precluded
the consideration of the forward walking as a stable and reliable
method. In contrast with these considerations, we find that these
statistical fluctuations (of unphysical origin) show a highly
depressed effect over integrated quantities, and that in order to
accurately sample pure estimators stable regions can be reached.
The method we present is somehow related to the one of Ref.
\onlinecite{BR91} but with the advantage of not requiring a
tagging algorithm.  As we shall show in Section 2, the averages
are basically taken as mixed expectation values, and therefore
the pure estimators can be readily incorporated in the original
Monte Carlo algorithms.

The layout of this paper is as follows.  In Section 2 the
algorithm for the evaluation of pure estimators of coordinate
operators is described.  In order to check the correctness of our
implementation in a Diffusion Monte Carlo code, as well as the
capability and resolution of the method, results for several
moments of H and H$_2$, where exact results are available, are
presented in Section 3. Pure results for the partial energies and
structure properties of liquid $^4$He are reported in Section 4.
The application of the method to a real many body problem is a
compelling test for its reliability and stability.  Finally, the
main conclusions are discussed in Section 5.

\section{Pure expectation values}

The pure estimator of an operator $A(\bR)$ (\ref{pure}) may be
written as
\beq
\langle A(\bR) \rangle_p=
\left\langle \Phi_0 (\bR) \, \left| \,
A(\bR)
\, \frac{\Phi_0(\bR)}{\psi(\bR)} \, \right | \, \psi(\bR) \right
\rangle \, \left/ \,
\left\langle \Phi_0 (\bR) \, \left| \,
 \frac{\Phi_0(\bR)}{\psi(\bR)} \, \right | \, \psi(\bR) \right
\rangle  \right.  \ .
\label{pure2}
\eeq
Following Liu {\it et al.}, \cite{LK74} $\Phi_0(\bR)/\psi(\bR)$
can be obtained from the asymptotic offspring of the $\bR$
walker.  In fact, assigning to each walker $\bR_i$ a weight
$W(\bR_i)$ proportional to its number of descendants
\beq
W(\bR)=n(\bR,t \rightarrow \infty) \ ,
\label{pes}
\eeq
Eq.  (\ref{pure2}) turns out to
\beq
\langle A(\bR)
\rangle_p=\frac{ \sum_i A(\bR_i) \, W(\bR_i)}{\sum_i W(\bR_i)} \ ,
\label{magia}
\eeq
where the summatory $\sum_i$ runs over all walkers and all times
in the asymptotic regime.  As it is clear from its proper
definition, the weight of a walker existing at time $t$, given by
Eq.  (\ref{pes}), is not known until a future time $t^{\prime}
\geq t+T$, being $T$ a time interval long enough so that Eq.
(\ref{pes}) could be replaced by $W(\bR(t))=n(\bR(t^{\prime}))$.

In order to proceed to the evaluation of Eq. (\ref{magia}) two
different approaches are possible.  In the first one, a tagging
algorithm capable of identifying, at any time, which walker of
any precedent configuration originated an actual walker could be
used. Then, one could determine the number of descendants of the
former $\bR_{i}$, and accumulate its contribution to Eq.
(\ref{magia}) ``from the distance". Such a tagging algorithm has
been devised in  Refs. \onlinecite{BR91,RN86,Runge}.  On the
other hand, one can work out an algorithm that operates with only
the actual values of $A(\bR_{i})$, in such a way that a weight
proportional to its future progeny is automatically introduced.
This second approach is the one we have followed in the present
work.

The schedule of the algorithm is the following. The set of
walkers at a given time $\{ \bR_i \}$ and the values that the
operator $A$ takes on them $\{A_i\}$ evolve, after a time step,
to
\beqn
\{\bR_i \} & \rightarrow & \{ \bR_i^{\prime} \}   \\[0.4cm]
\{A_i \} & \rightarrow & \{ A_i^{\prime} \} \ .
\eeqn
In the same time interval, the number of walkers $N$ changes to
$N^{\prime}$. In order to sample the pure estimator of $A$, we
introduce an auxiliary variable $\{P_i\}$, associated to each
walker, with an evolution law given by
\beq
\{P_i \} \rightarrow \{P_i^{\prime} \} = \{ A_i^{\prime} \}+\{P_i^t \} \
,
\label{patum}
\eeq
where $\{P_i^t\}$ is the old set $\{P_i\}$ ``transported" to the
new one, in the sense that each element $P_i$ is replicated as
many times as the $\bR_i$ walker, without any other changes. $\{
P_i \}$ is initialized to zero when the run starts.

With this procedure, after $M$ addition steps (\ref{patum}) we
end up with a set of $N_f$ values $\{P_i\}$. A pure estimator of
$A$ is given by
\beq
\langle A(\bR) \rangle_p = \sum_{i=1}^{N_f} \{P_i\}\, / \, (M \times
N_f) \ .
\label{magic2}
\eeq

The contribution to the $\{ P_i \}$  entering in Eq.
(\ref{magic2}), coming from a generic walker $\bR(t)$, can be
determined following the evolution of the series. Clearly there
is no contribution from $\bR(t)$ until time $t$ is reached. At
this moment, $A (\bR(t))$ enters in one of the rows of $\{P\}$
(\ref{patum}).  From now on, if any of the descendants of
$\bR(t)$ disappears or replicates, the former contribution so
does. As a result, $A(\bR(t))$ appears in as many rows of $\{P\}$
as descendants of $\bR(t)$ exist, and therefore its contribution
to Eq. (\ref{magic2}) is proportional to the weight
$W(\bR(t))$(\ref{pes}).

A final regard concerning the implementation of the algorithm has
to be made. In Eq. (\ref{patum}), the ``transport" operation
accounts for the replication of the $A(\bR)$ contribution. In
order to ensure the asymptotic condition (\ref{pes}), the series
are continued for a while only with the reweighting law
\beq
\{P_i \} \rightarrow \{ P_i^{\prime} \}=\{P_i^t\} \ .
\label{transport}
\eeq
Since a calculation is usually divided in several blocks, one can
collect data during a block and allow for a further reweighting
in the following one. In this second block, new information can
be accumulated to be reweighted in the next block.  This
mechanism can be incorporated in the algorithm in a rather simple
way. The final result is that, after a first initialization
block, each new block gives a value for the pure expectation
value of $A$.

An alternative to the simple branching algorithm, implicitly
assumed in the above method, is the use of weights $p(\bR_i)$
related to the branching factor. In fact, it has been proved
\cite{BR91,Runge} that the branching algorithms with weighting
allow for some reduction in the variance of the expectation
values. Our method for computing pure estimators is easily
extended to these algorithms. In particular, the evolution laws
(\ref{patum},\ref{transport}) become:
\beqn
\{P_i \} & \rightarrow & \{P_i^{\prime} \} = \{
p(\bR_i^{\prime}) \times A_i^{\prime} \}+ \left\{
\frac{p(\bR_i^{\prime})}{p(\bR_i)} \times P_i^t \right\}  \\[0.2cm]
\{P_i \} & \rightarrow & \{ P_i^{\prime}
\}=\left\{ \frac{p(\bR_i^{\prime}) }{p(\bR_i)} \times P_i^t \right\} \ ,
\label{refino}
\eeqn
whereas the expression of the pure expectation value
(\ref{magic2}) is only modified by a normalization factor.

\section{Application to simple systems: H and H$_2$}
As a test of the algorithm developed in the preceding Section, we
present results for the lowest coordinate moments of the hydrogen
atom H and the hydrogen molecule H$_2$. In both systems the
nuclei are kept fixed and relativistic corrections are neglected.
Atomic units have been used in all the Section.

The DMC program used in the calculations is exact up to order
$(\Delta t)^2$ in the short--time approximation for the Green's
function. More specific details of the algorithm are given in
Ref. \onlinecite{BoCas}.  Chin
\cite{Chin} has extensively discussed the relation between a quadratic
time--step dependence of the eigenvalue and the violation of the
cusp condition in electronic systems. However, the achievement of
a quadratic dependence in $\Delta t$, which has also been
discussed by Umrigar {\em et al.}, \cite{Umrigar} is not the main
objective of the present work. Our aim is to check the efficiency
of the algorithm for the extraction of pure expectation values in
simple systems, as H and H$_2$, where exact results are
available.

We have tested the reliability of our method including it in two
versions of the original code, corresponding to the use or not of
the weights $p(\bR_{i})$ (15,16). In both cases, satisfactory
results are obtained, the variance of the mixed and pure
estimators being slightly reduced when the weights $p(\bR_{i})$
are considered in the branching process.

\subsection{Hydrogen atom}
Two different guiding functions are used for importance sampling
in the H calculation. The first one corresponds to a $1s$ Slater
orbital
\beq
\psi_I=\exp (- \alpha r)
\label{psi1h}
\eeq
with a slightly modified exponent ($\alpha=0.9$). The second one
is taken as the product of the $1s$ Slater orbital by a gaussian
\beq
\psi_{II}=\exp (- \alpha r-\beta r^2)
\label{psi2h}
\eeq
with $\alpha=1.0$ and $\beta=0.06$. The analytic variational
energies are $E_v=-0.495$ for $\psi_I$ and $E_v=-0.4853$ for
$\psi_{II}$, to be compared with the exact result $E=-0.5$. A
difference between $\psi_I$ and $\psi_{II}$, which could be
relevant in the time--step dependence of the energy, is that
whereas $\psi_{II}$ satisfies the cusp condition $\psi_I$ does
not.  As far as the DMC calculation is concerned, we have used
$\Delta t=0.05$ in both cases with no significant differences
when $\Delta t$ is reduced by a factor of two. The number of
walkers was maintained at a value of 700 with an unnoticeable
bias respect to larger populations. The samplings were performed
over approximately $4 \cdot 10^5$ configurations.

In Table I variational (\ref{variatio}), mixed (\ref{mixed}) and
extrapolated (\ref{extrapol}) estimators of the potential energy
$V$, the radial distance $r$, the squared radial distance $r^2$
and $z^2$ are reported in comparison with the exact results. The
extrapolated expectation values improve the mixed results lying
near the exact ones.  However, some differences which depend on
the trial function used for importance sampling remain, showing
that the extrapolation method suffers from a systematic bias
related to $\psi$.

The pure expectation values of the same coordinate moments are
reported in Table II.  Neither the $\psi_I$ nor the $\psi_{II}$
results are biased with respect to the exact values. In fact, as
it happens in the exact mixed estimator for the Hamiltonian, the
quality of the trial function is only reflected in the magnitude
of the variance.  This influence may be observed in the larger
errors of the pure estimators for $\psi_{II}$ with respect to the
ones for $\psi_I$.

The DMC calculation is divided in blocks of a number of
iterations $\Delta L$. According to the algorithm developed in
Section 2 the block length has to be long enough to ensure the
pure estimation in the asymptotic regime. In Figs. 1 and 2, the
$\Delta L$--dependence of the pure expectation value for $r^2$ is
plotted for $\psi_I$ and $\psi_{II}$, respectively. Also shown
are the exact result (solid line) and the extrapolated estimator
(dashed line) corresponding to the trial function used in the
calculation. The bias coming from the wave function components
other than the ground state is rapidly suppressed, as expected
from the evolution law $\exp (- H t)$.  The asymptotic condition
is satisfied in both cases for values $\Delta L \geq 500$.
Beyond a transient regime, the prediction of the pure estimator
is stable for a wide range of $\Delta L$ values with a negligible
systematic bias. The statistical error in the $\psi_{II}$ case is
larger than in the $\psi_I$ one but, in both cases, the central
value reproduces accurately the exact results. On the other hand,
the extrapolated predictions are biased respect to the exact and
pure values, significantly for $\psi_{II}$ as expected from its
poorer variational quality.

\subsection{Hydrogen molecule}
The trial wave function we have used in the study of the hydrogen
molecule is of the form \cite{RC82}
\beq
\Phi = \phi(r_{1A},r_{1B}) \, \phi(r_{2A},r_{2B}) \, \exp (a
r_{12}/(1+b r_{12}) )
\label{phimol}
\eeq
with the molecular orbital
\beq
  \phi(r_{iA},r_{iB}) = \exp (-\zeta \, r_{iA}) + \exp ( -\zeta
\, r_{iB}) \ .
\label{phiorb}
\eeq
The distances $r_{iA}$, $r_{iB}$ correspond to the
electron--nucleus separation, and $r_{12}$ stands for the
electron--electron distance. The internuclear separation is kept
fixed in the equilibrium distance $r_{AB}=1.401$. In order to
test the accuracy of the pure algorithm, we have used the trial
function (\ref{phimol}--\ref{phiorb}) with two different sets of
parameters: \cite{Chin}
\beq
\psi_I= \Phi(\zeta=1.189,\ a=0.50,\ b=0.40)
\label{phi1mol}
\eeq
and
\beq
\psi_{II}= \Phi(\zeta=1.189,\ a=0.0,\  b=0.0) \ .
\label{phi2mol}
\eeq

The value of $\zeta$ is obtained from the cusp condition between
an electron and a nucleus ($\zeta=1+\exp (-\zeta \, r_{AB})$).
The Jastrow factor between the electrons appearing in the general
form (\ref{phiorb}) is suppressed in $\psi_{II}$ whereas it is
considered in $\psi_I$ with a value for the parameter $a$ which
guarantees the electronic cusp condition. The variational
energies are $E_v=-1.1471(9)$ for $\psi_I$ and $E_v=-1.1288(8)$
for $\psi_{II}$, to be compared with the exact result $E=-1.174
\, 47 \ldots$.
\cite{Kolos}

The DMC calculations have been carried out with $\Delta t=0.05$
and an average population of 500 walkers.  The sampling has been
made over $2 \cdot 10^5$ configurations. We have checked that the
biases due to the time step and the finite population are smaller
than the statistical error.

The H$_2$ results closely follow the trends observed in the H
calculation.  The $\Delta L$--dependence in the pure estimator of
$r^2$ is shown in Fig. 3 for $\psi_I$, and in Fig. 4 for
$\psi_{II}$. Similar behaviours have been obtained for the other
operators. One can see that the asymptotic region is already
reached at relatively small values of $\Delta L$. In the
asymptotic regime, the pure values fluctuate around the exact
value \cite{Kolos2} (solid line) with a statistical noise related
to  the quality of the trial function used for importance
sampling. The growth of the error bars due to the death of
walkers in the forward walking process is only significant for
the largest $\Delta L$ values.  The extrapolated prediction
(dashed line) is manifestly biased.

In Table III, we report results for variational, mixed and
extrapolated estimators of $V$, $r^2$ and $z^2$ using $\psi_I$
(\ref{phi1mol}) and $\psi_{II}$ (\ref{phi2mol}) as trial
functions.  As one can see, the variational results  of the
coordinate moments for $\psi_{II}$ are closer to the exact values
than for $\psi_I$, although $\psi_I$ is energetically preferred
to $\psi_{II}$. Nevertheless, the simple extrapolated expectation
values are in neither case statistically compatible with the
exact values. \cite{Kolos2} In Table IV the pure expectation
values for the same moments are reported in comparison with the
exact results. In spite of the shortcomings of these trial
functions in reproducing the properties of H$_2$, the pure
estimator does reproduce the exact values independently of the
trial wave function.

\section{Liquid $^{\bf 4}$He}
Domain GFMC and DMC have been extensively applied to the study of
the ground state properties of liquid $^4$He for the last fifteen
years.
\cite{BoCas,Kalee} The
exactness of these methods joined with the accuracy in the
knowledge of the He interatomic potential have made feasible an
excellent agreement between theoretical results and experimental
data. In order to sample expectation values other than the
Hamiltonian, e.g., partial energies or the radial distribution
function $g(r)$, the extrapolation method has been commonly used.
In spite of the success in describing properties as $g(r)$ a
small bias is present in integrated quantities as the potential
energy. Furthermore, the extrapolated estimator has evidenced its
shortcomings in the calculation of density profiles of $^4$He
clusters producing unphysical negative values for $\rho(r)$ in
the cluster surface. \cite{Chinkro} We have applied the algorithm
developed in Section 2 to bulk liquid $^4$He in order to show
both its applicability to a fully many body problem and its
capacity of removing the uncertainties introduced by the
extrapolation method.

The Schr\"odinger equation is solved by means of a Quadratic
Diffusion Monte Carlo method considering the $N$--particle
Hamiltonian
\beq
H=-\frac{\hbar^2}{2 \, m} \, \nabla_{\bf R}^2 + V({\bf R})\ ,
\eeq
being ${\bf R}=({\bf r}_1,\ldots,{\bf r}_N)$ and $V({\bf R})$ the
interatomic potential. The results presented below have been
obtained considering the HFD--B(HE) potential proposed by Aziz
{\it et al.}
\cite{AzizII}
In a previous paper \cite{BoCas,BoCasII} we have shown that this
renewed version of the well--known Aziz potential \cite{Azizvell}
improves appreciably the equation of state of liquid $^4$He with
respect to the Aziz results.

In order to establish the influence of the trial wave function
used for importance sampling several options have been
considered. The first one is the simple McMillan two--body
function \cite{McM}
\beq
\psi_{J1}=\prod_{i<j} \exp \left[ - \frac{1}{2} \, \left(
\frac{b}{r_{ij}} \right)^5 \right]
\label{mcmillan}
\eeq
with $b=1.20\ \sigma$ ($\sigma=2.556\ $\AA). The second one is an
improved version of (\ref{mcmillan}) proposed by Reatto
\cite{Reatto}
\beq
\psi_{J2}=\prod_{i<j}\exp \left\{ -\frac{1}{2} \, \left(
\frac{b}{r_{ij}} \right)^5 - \frac{L}{2} \, \exp \left[ - \left(
\frac{r_{ij}-\lambda}{\Lambda} \right)^2 \right] \right\}
\label{reatto}
\eeq
with $b=1.20\ \sigma$, $L=0.2$, $\lambda=2.0 \ \sigma$, and
$\Lambda=0.6\ \sigma$. Finally, we have also used a trial wave
function which contains three--body correlations \cite{Schmidt}
\beq
\psi_{JT}=  \psi_{J1} \ \exp \left[ -\frac{1}{4} \lambda \sum_k {\bf
G}_k \cdot {\bf G}_k + \frac{1}{2} \lambda \sum_{i<j}
\xi^2(r_{ij}) \, r_{ij}^2 \right]
\eeq
where
\beq
{\bf G}_k=\sum_{l \neq k} \xi(r_{kl}) {\bf r}_{kl} \ ,
\eeq
and
\beq
\xi(r)=\exp\left[ - \left( \frac{r-r_t}{r_{\omega}} \right)^2 \right] \
{}.
\eeq
The values for the triplet parameters are $\lambda=-1.08\
\sigma^{-2}$, $r_t=0.80\ \sigma$ and $r_{\omega}=0.41\ \sigma$.
In the three trial functions, the values of the parameters are
those which optimize the variational energy at the experimental
equilibrium density $\rho_0^{\rm expt}=0.365\
\sigma^{-3}$. All the results presented below correspond to the
density $\rho_0^{\rm expt}$ that coincides with the theoretical
equilibrium density. \cite{BoCas}

Results for the potential energy per particle using $\psi_{J1}$,
$\psi_{J2}$ and $\psi_{JT}$ as importance sampling are reported
in Table V. A small but significant difference between the
extrapolated results $\langle V/N \rangle_e$ appear, pointing to
a bias related to the quality of the trial wave function. The
bias is completely removed when the pure estimator is calculated,
as one can see in the last row of Table V. The three values for
$\langle V/N \rangle_p$ are indistinguishable and, what is more
important, they evidence a systematic error of the extrapolation
approximation. In fact, none of the extrapolated values is
statistically compatible with the common pure value, being the
closest estimation the one obtained with $\psi_{JT}$ which
actually is the best variational choice. Considering the result
for the energy per particle $E/N=-7.267 \pm 0.013$ K,
\cite{BoCas} the pure result for the kinetic energy is $T/N=14.32
\pm 0.05$ K. Experimental determinations from analysis of deep
inelastic scattering data predicts a slightly lower value
$(T/N)^{\rm expt}=13.3 \pm 1.3$ K, \cite{Sokol} being the
difference mainly due to the significant errors in the
experimental measurement of the tail of the response function.

As far as the stability of the method is concerned, the
dependence of the pure expectation value of $V/N$ on the length
of the forward walking is plotted in Fig. 5. The results obtained
(points with errorbars) follow the trends observed in Section 3
(Figs. 1--4). After a transition regime, and already for
relatively small $\Delta L$ values ($\Delta L \geq 250$), an
asymptotic limit is reached where the systematic error is
practically negligible. Notice that in the simple algorithm we
have presented in Section 2 a forward walking of length $\Delta
L$ is constructed from data ranging from $L$ to $2L$, and hence
the length of the forward walking is not the same for all the
walkers.  This effect is not relevant provided that a region of
stability exists.  On the other hand, one can determine the
asymptotic value within a single run collecting data for
different block lengths. The statistical errors associated to
each individual $\Delta L$ value can be lowered in the same way
as mixed estimators, i.e., continuing the evolution of the
series. The biases associated to the extrapolated expectation
values are also shown in Fig. 5, where $\langle V/N\rangle_e$
using $\psi_{J2}$ and $\psi_{JT}$ are represented by a
long--dashed and a short--dashed lines, respectively.

Other important quantities in the study of quantum liquids can
also be calculated with the pure algorithm. In particular, the
two--body radial distribution function
\beq
g(r_{12})=\frac{N\,(N-1)}{\rho^2}\
\frac{\int |\Phi_0({\bf
r}_1,\ldots,{\bf r}_N) |^2\, d{\bf r}_3 \ldots d{\bf r}_N} {\int
|\Phi_0({\bf r}_1,\ldots,{\bf r}_N ) |^2\, d{\bf r}_1 \ldots
d{\bf r}_N}
\label{ger}
\eeq
and the static structure function
\beq
S(q) = \frac{1}{N}\ \frac{\langle \Phi_0 | \rho_{\bf q} \,
\rho_{-{\bf q}} | \Phi_0 \rangle}{\langle \Phi_0 | \Phi_0
\rangle} \ ,
\label{seq}
\eeq
with
\beq
 \rho_{\bf q} = \sum_{i=1}^{N} e^{i {\bf q} \cdot {\bf r}_i } \ .
\label{densi}
\eeq
The result obtained for $g(r)$ is shown in Fig. 6 in comparison
with the experimental data of Ref. \onlinecite{Sven}. As one can
see, the pure expectation value of $g(r)$ is in a good agreement
with the experimental $g(r)$ for all the calculated $r$ values.
In Fig. 7 the pure structure function $S(q)$ is plotted together
with the experimental measures of Refs.
\onlinecite{Sven,Hallock}. An overall agreement between the theoretical
and experimental $S(q)$ is obtained, lying our result well
between the two experimental determinations.  The extrapolated
estimations of $g(r)$ and $S(q)$ \cite{BoCas,Kalee} are not
significantly different of the pure result. It is clear that the
difference between the results provided by the extrapolated and
pure estimators is larger for integrated quantities as, for
instance, the partial energies.

 \section{Concluding remarks} We have presented a simplified
version of the forward walking method to obtain unbiased
expectation values for operators that do not commute with the
Hamiltonian. One of the advantages of this algorithm, in front of
others based on side walks or bilinear sampling, is that it
enters ``naturally" in any domain or short--time Green's function
Monte Carlo program. The sampling of the pure estimator closely
follows the standard procedure to sample mixed expectation
values. Nevertheless, the main point is the stability of the
method. The exponential decrease of the bias with the forward
walking length and the evidence of a slow increase of the
statistical uncertainties for {\em physical} quantities result in
stability regions large enough to sample pure expectation values
with negligible biases. In contrast, we do find also the common
result that the weight associated with the offspring of an
individual walker fluctuates all along the stability region.
However, looking for stability in the values of the weights is
more than what should be asked for. What is mainly required from
a Monte Carlo method is stability in the expectation value of an
operator $A$, $\langle A \rangle_p$. The computation of this
average has clearly a much better chance of success.

The accuracy of the method has been first verified in the H atom
and the H$_2$ molecule.  In both systems, the pure expectation
values reproduce the exact results with statistical errors which
are not appreciably larger than the ones associated to the
extrapolated predictions.  In all cases, the extrapolated
expectation values appear significantly biased respect to the
pure/exact values in a quantity which is related to the trial
function but difficult to assert {\em a priori}.

Finally, the pure algorithm has been applied to study some
properties of liquid $^4$He at zero temperature. The
implementation of this algorithm in a many body problem is also
quite straightforward and the results obtained follow the same
trends analysed in the simple systems (H,H$_2$). The method is
stable and generates results which are not biased by the
importance sampling as it happens with the extrapolated
estimations. In order to reduce the error bars of the pure values
to the level of those associated to the total energy, the series
have to be a bit longer. However, one does not have to perform
the auxiliary Variational Monte Carlo calculation required by the
extrapolation method and, more importantly, the guarantee of an
exact result is fulfilled.

\acknowledgments
This work has been supported in part by DGICYT (Spain) Grant Nos.
PB92--0761 and PB90--06131.

%**FIGURES

\begin{figure}

\caption{Pure expectation value of $ r^2 $ for H as a function of
the block length $\Delta L$ using $\psi_I$. The solid and dashed
lines correspond to the exact and extrapolated results,
respectively.}

\end{figure}

\begin{figure}

\caption{Same as in Fig. 2 but for
$\psi_{II}$.}

\end{figure}

\begin{figure}

\caption{Pure expectation value of $ r^2 $ for H$_2$ as a
function of the block length $\Delta L$ using $\psi_I$.  The
solid and dashed lines correspond to the exact and extrapolated
results, respectively.}

\end{figure}

\begin{figure}

\caption{Same as in Fig. 4 but for $\psi_{II}$.}

\end{figure}

\begin{figure}

\caption{Pure expectation value of $V/N$ for liquid $^4$He at
$\rho_0^{\rm expt}$ as a function of the block length $\Delta L$.
The long--dashed and short--dashed lines stand for the
extrapolated estimations using $\psi_{J2}$ and $\psi_{JT}$,
respectively.}

\end{figure}

\begin{figure}

\caption{Pure expectation result of the two--body radial
distribution function (solid line)
for liquid $^4$He at $\rho_0^{\rm expt}$
in comparison with the experimental data (points) of
Ref. \protect\onlinecite{Sven}.}

\end{figure}

\begin{figure}

\caption{Pure estimation of the static structure function
(points) for liquid $^4$He at $\rho_0^{\rm expt}$ in comparison
with the experimental determinations of Refs.
\protect\onlinecite{Sven} (solid line) and
\protect\onlinecite{Hallock} (dashed line). The error bars of the
theoretical points are only depicted where larger than the size
of the symbols.}

\end{figure}

%***TABLES

\begin{table}

\caption{Variational, mixed and extrapolated expectation values
for H using $\psi_I$ and $\psi_{II}$. All the results are
analytic.}

\begin{tabular}{lllll}
 {}  & \ \ \ $\langle V \rangle$ &  \ \ $\langle r \rangle$  &  \
\ $\langle r^2 \rangle$ & \ \ $\langle z^2 \rangle$   \\ \tableline
 $\psi_I$     &           &        &        &        \\
 Variational  &  -0.9000  & 1.6667 & 3.7037 & 1.2346 \\
 Mixed  &  -0.9500  & 1.5789 & 3.3241 & 1.1080 \\
 Extrapolated &  -1.0000  & 1.4912 & 2.9445 & 0.9815 \\
 $\psi_{II}$  &           &        &        &        \\
 Variational  &  -1.1507  & 1.2560 & 2.0333 & 0.6778 \\
 Mixed& -1.0818  & 1.3623 & 2.3072 & 0.7690 \\
 Extrapolated  & -1.0129  & 1.4686 & 2.5811 & 0.8602 \\
 Exact         & -1.0     & 1.5    & 3.0    & 1.0     \\
\end{tabular}

\end{table}

\begin{table}

\caption{Pure expectation values for H, using $\psi_I$ and
$\psi_{II}$, in comparison with the exact results. The
statistical errors are indicated in parentheses.}

\begin{tabular}{lllll}
 {}  & \ \ \ $\langle V \rangle$ &  \ \ $\langle r \rangle$  &  \
\ $\langle r^2 \rangle$ & \ \ $\langle z^2 \rangle$
 \\ \tableline
 $\psi_I$     & -0.9987(10) & 1.4999(10) & 3.0002(36) & 1.0004(17)
\\
 $\psi_{II}$  & -0.9975(14) & 1.4993(28) & 2.995(14) & 0.9964(61)
\\
 Exact         & -1.0     & 1.5    & 3.0    & 1.0     \\
\end{tabular}

\end{table}

\begin{table}

\caption{Variational, mixed and extrapolated
expectation values
for H$_2$ using $\psi_I$ and $\psi_{II}$.}

\begin{tabular}{llll}
 {}  & \ \ \ $\langle V \rangle$ &  \ \ $\langle r^2
\rangle$ & \ \ $\langle z^2 \rangle$   \\ \tableline
 $\psi_I$     &           &        &             \\
 Variational  &  -2.1034(9)  & 3.0740(1)  & 1.2813(38) \\
 Mixed        &  -2.2415(7)  & 2.7530(14) & 1.1281(12) \\
 Extrapolated &  -2.3796(17) & 2.4320(28) & 0.9749(45) \\
 $\psi_{II}$  &           &        &             \\
 Variational  &  -2.2254(19) & 2.6155(23) & 1.0942(22) \\
 Mixed        &  -2.3012(13) & 2.5416(42) & 1.0409(19) \\
 Extrapolated  & -2.3770(32) & 2.4677(87) & 0.9876(44) \\
 Exact         & -2.3489     & 2.5464     & 1.0230     \\
\end{tabular}

\end{table}

\begin{table}

\caption{Pure expectation values for H$_2$,
using $\psi_I$
and $\psi_{II}$, in comparison with the exact results.}

\begin{tabular}{llll}
 {}  & \ \ \ $\langle V \rangle$ &  \ \ $\langle r^2
\rangle$ & \ \ $\langle z^2 \rangle$   \\ \tableline
 $\psi_I$     & -2.3448(24) & 2.5424(46) & 1.0244(23) \\
 $\psi_{II}$  & -2.3454(39) & 2.5412(74) & 1.0210(44) \\
 Exact        & -2.3489     & 2.5464     & 1.0230     \\
\end{tabular}

\end{table}

\begin{table}

\caption{ Variational, mixed, extrapolated and pure
expectation values of $V/N$ (in K) for liquid $^4$He at
$\rho_0^{\rm expt}$ using different trial wave functions $\psi$.}

\begin{tabular}{llll}
{}  & \ \ \ \ $\psi_{J1}$ &  \ \ \ \ $\psi_{J2}$  &  \ \ \ \
$\psi_{JT}$
\\ \tableline
$\langle V/N \rangle _v$  &  -21.054(26) &  -21.311(18)
& -21.348(20) \\
$\langle V/N \rangle_m$  &  -21.459(8)  &  -21.600(8)
& -21.541(8) \\
$\langle V/N \rangle _e$  &  -21.864(30)  &  -21.889(24)
& -21.734(25) \\
$\langle V/N \rangle _p$  &  -21.56(5)  &  -21.59(5)  &
-21.58(5) \\
\end{tabular}

\end{table}

\end{document}